\documentclass[aps,prd,secnumarabic,nobibnotes,twocolumn,superscriptaddress]{revtex4-1}
\usepackage{amsfonts}
\usepackage{mathrsfs}
\usepackage{amsmath}
\usepackage{color}
\usepackage{natbib}
\usepackage{graphicx}
\usepackage{bm}
\usepackage{amssymb}
\usepackage{xspace}
\usepackage{epstopdf}
\usepackage{dcolumn}
\usepackage{multirow}
\usepackage[colorlinks=true, letterpaper=true, pdfstartview=FitV, linkcolor=blue, citecolor=blue, urlcolor=blue]{hyperref}
\usepackage{wrapfig}

\makeatletter

\newcommand{\Rmnum}[1]{\expandafter\@slowromancap\romannumeral #1@}
\makeatother

\begin{document}
\title{Single pair of type-\uppercase\expandafter{\romannumeral3} Weyl points half-metals: BaNiIO$_6$ as an example}

\author{Guangqian Ding}\thanks{These authors contributed equally to this work.}
\address{School of Science, Chongqing University of Posts and Telecommunications, Chongqing 400065, China}

\author{Jianhua Wang}\thanks{These authors contributed equally to this work.}
\address{School of Physical Science and Technology, Southwest University, Chongqing 400715, China}

\author{Zhi-Ming Yu}
\address{Centre for Quantum Physics, Key Laboratory of Advanced Optoelectronic Quantum Architecture and Measurement (MOE), School of Physics, Beijing Institute of Technology, Beijing, 100081, China}
\address{Beijing Key Lab of Nanophotonics $\&$ Ultrafine Optoelectronic Systems, School of Physics, Beijing Institute of Technology, Beijing, 100081, China}

\author{Zeying Zhang}\thanks{Corresponding authors}\email{zzy@mail.buct.edu.cn}
\address{College of Mathematics and Physics, Beijing University of Chemical Technology, Beijing 100029, China}

\author{Wenhong Wang}
\address{Tiangong University, Tianjin 300387, China}

\author{Xiaotian Wang}\thanks{Corresponding authors}\email{xiaotianwang@swu.edu.cn and xiaotianw@uow.edu.au}
\address{School of Physical Science and Technology, Southwest University, Chongqing 400715, China}
\address{Institute for Superconducting and Electronic Materials (ISEM), University of Wollongong, Wollongong 2500, Australia}

\begin{abstract}
The realization of Weyl systems with the minimum nonzero number of Weyl points (WPs) and \textit{full} spin polarization remains challenging in topology physics and spintronic. In this study, \textit{for the first time}, we used first-principle calculations and symmetry analysis to demonstrate that BaNiIO$_6$, a dynamically and thermodynamically stable half-metallic material, hosts fully spin-polarized single-pair WPs (SP-WPs) with a charge number ($\cal{C}$) of $\pm$2 and a type-\uppercase\expandafter{\romannumeral3} band dispersion around the Fermi level. Moreover, the fully spin-polarized SP-WPs induce double-helicoid Fermi arcs on the (10$\overline{1}$0) surface. The half-metallic state and the spin-polarized SP-WPs are robust to uniform strains (from -10\% to +8\%) and on-site Hubbard-Coulomb interactions (from 0 eV to 6 eV). When +9\% or +10\% uniform strain is applied to the BaNiIO$_6$ system, it hosts six additional type-\uppercase\expandafter{\romannumeral2} WPs with $\lvert{\cal{C}}\rvert=1$ in the three-dimensional Brillouin zone in addition to the two type-\uppercase\expandafter{\romannumeral3} WPs with $\lvert{\cal{C}}\rvert=2$. We hope that this study will motivate future research into SP-WPs half-metals.
\end{abstract}
\maketitle

\section{Introduction}
Weyl semimetals~\cite{add1,add2,add3,add4,add5} are crystals that host Weyl points (WPs) between two bands and have characteristic surface Fermi arcs in their electronic structure. Weyl semimetals can appear in solids when a material breaks either inversion symmetry~\cite{add6,add7,add8,add9,add10,add11}, time-reversal symmetry~\cite{add1,add4,add12,add13}, or both~\cite{add14}. It would be highly preferable to realize Weyl semimetals with a minimum number of WPs in order to have a clean system to carefully study the properties of Weyl fermions. A minimum number of WPs will simplify the theoretical model and transport experiments~\cite{add15}. Moreover, the surface states that arise from the minimum number of WPs can be easily imaged in spectroscopy experiments. However, if other crystal symmetry acts on the materials (i.e., rotates or reflects them), there can be many additional WPs in the systems; for example, there are 24 WPs in TaAs, TaP, NbAs, and NbP~\cite{add6,add7,add8,add9,add16,add17,add18,add19}, 8 in MoTe$_2$~\cite{add20}, 8 in chalcopyrites CuTlSe$_2$, AgTlTe$_2$, AuTlTe$_2$, and ZnPbAs$_2$~\cite{add21}, and 60 in the stoichiometric compound SrSi$_2$~\cite{add22}.

The well-known arguments are that four (two) is the minimum number of WPs allowed in a \textit{nonmagnetic} (\textit{magnetic}) Weyl semimetal~\cite{add1,add12,add15,add23,add24,add25,add27,add28}. However, these arguments are flawed, as demonstrated by our recent study~\cite{add29}, which revealed that nonmagnetic spinless systems can have only two WPs located at two high-symmetry time-reversal-invariant momentum (TRIM). However, the minimum number of WPs at the TRIM for spinful systems should be eight because of the Kramers degeneracy at all eight TRIM points in the three-dimensional (3D) Brillouin zone (BZ).

\textit{An interesting question is whether an electronic system can have ferromagnetism, 100\% spin polarization, and only two WPs at TRIM points.} In this work, we answer this question affirmatively. Herein, for the first time, we propose single-pair WPs (SP-WPs) half-metals, a new type of system that provides a new simple platform for the interplay between magnetism and topological behaviors. The term ``SP" refers to the minimal number of WPs (i.e., two) that should appear in a topological crystal in accordance with the no-go theorem~\cite{add30,add31}. Moreover, in this study, we used a dynamically and mechanically stable material, BaNiIO$_6$ with space group $P312$ [No. 149], as an example to prove the theory behind half-metallism, ferromagnetism, and SP-WPs at TRIM points.

\begin{figure}
\includegraphics[width=8.5cm]{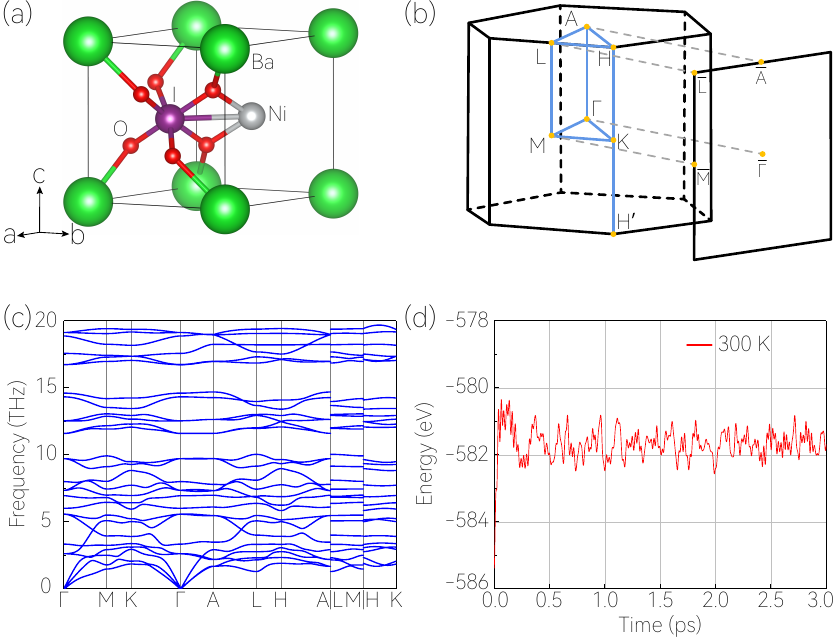}
\caption{(a) Crystal structure of the BaNiIO$_6$ unit cell. (b) 3D bulk and two-dimensional (2D) (10$\overline{1}$0) surface BZs. (c) Phonon dispersion of BaNiIO$_6$. A 2 $\times$ 2 $\times$ 2 supercell is adopted for the calculation of force constants. (d) Total energy fluctuation of a 3 $\times$ 2 $\times$ 2 BaNiIO$_6$ supercell during ab initio molecular dynamics (AIMD) simulations at 300 K.
\label{fig1}}
\end{figure}

Interestingly, our theoretical simulations on the spin-polarized band structures of BaNiIO$_6$ demonstrated that its WPs are unique in the following aspects. (1) There are only two WPs around the Fermi level ($E_F$) in the spin-up channel: a WP with a charge number ($\cal{C}$) of 2 and another WP with  ${\cal{C}}= -2$, both of which are located at TRIM points $A$ and $\Gamma$, respectively. The large separation between the two WPs (at $A$ and $\Gamma$) allows the physics of the WPs to be exposed without intervention. (2) When spin-orbit coupling (SOC) is neglected, the SP-WPs proposed herein are completely formed by bands from the spin-up channel, making them 100\% spin-polarized. The fully spin polarized fermions~\cite{add32,add32a,add32b,add33,add34,add35,add36,add37,add38,add39,add40} may be helpful for spintronics applications. (3) The SP-WPs have a type-\uppercase\expandafter{\romannumeral3} band dispersion~\cite{add41,add42,add43,add44}, implying that they host a unique Fermi surface consisting of two touching electron-like pockets. (4) The SP-WPs are robust to uniform strains and on-site Hubbard-Coulomb interactions.

\section{Material example: BaNiIO$_6$}
BaNiIO$_6$ is expected to crystallize in the trigonal $P312$ space group [No. 149]. The experimental investigation of BaNiIO$_6$ may refer to that of the same mixed-metal periodates family~\cite{add44a,add44b}. Figure~\ref{fig1}(a) depicts the crystal structure of BaNiIO$_6$ retrieved from the Materials Project database~\cite{add45}. The derived lattice constants with the help of first-principle calculations of BaNiIO$_6$ are $a = b = 5.186$ {\r{A}} and $c = 5.857$ {\r{A}}. More details about the computational methods can be found in Supplemental Material (SM)~\cite{add46}. Ba, Ni, I, and O are located at the $1a$, $1d$, $1f$, and $6l$ Wyckoff sites in the structure, respectively. Table~\ref{Tab1} shows that almost all the magnetic moments are localized around the Ni atoms, indicating that the Ni atoms are primarily responsible for the magnetic nature of BaNiIO$_6$. The atomic magnetic moments listed in Table~\ref{Tab1} are computed by placing a sphere around each atom with the radius given in Table S1 (see SM~\cite{add46}). Different magnetic configurations in the 1 $\times$ 1 $\times$ 2 and 2 $\times$ 2 $\times$ 1 supercells, including ferromagnetic, antiferromagnetic, and nonmagnetic configurations, were considered to examine the magnetic ground state of BaNiIO$_6$ (see Figs. S1 and S2 in SM~\cite{add46}). The ferromagnetic state was discovered to have the lowest energy among all the magnetic configurations. In the following discussion, we only focus on the electronic structure and related topological signature of BaNiIO$_6$ with a ferromagnetic configuration.

\begin{figure*}
\includegraphics[width=13cm]{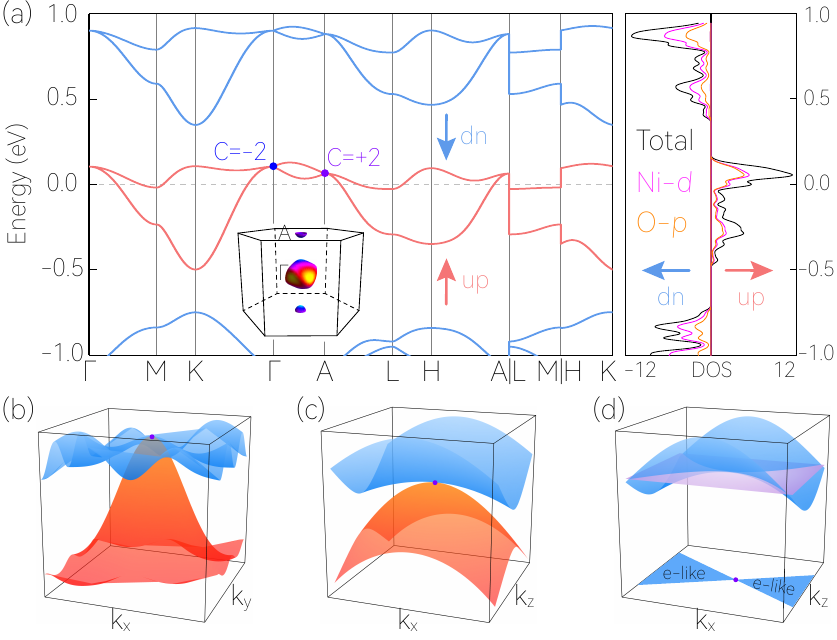}
\caption{(a) Spin-polarized band structures and density of states for the BaNiIO$_6$ ferromagnetic half-metal. The dn and up represent the spin-down and spin-up directions, respectively. The inset is a 3D Fermi surface at the $E_F$. Note that the density of states for Ba and I are not plotted because they are zero around the Fermi level. (b) The 3D plot of the spin-up bands around the WP at $A$ in the $k_x$-$k_y$ plane. (c) The 3D plot of the spin-up bands around the WP at $A$ in the $k_x$-$k_z$ plane. (d) Constant energy contour in the $k_x$-$k_z$ plane at the energy of the WP at $A$. The blue regions in (d) represent the connected electron-like (e-like) pockets.
\label{fig2}}
\end{figure*}

\begin{table}
\renewcommand\arraystretch{1}
\caption{\label{Tab1}Total and atomic magnetic moments ($\mu_B$) in the BaNiIO$_6$ unit cell.}
\begin{ruledtabular}
\begin{tabular}{ccccc}
       Total & Ba  & Ni  & I & O \tabularnewline
       \hline
    0.9991 & 0.007 & 0.742 & 0.03 & 0.032 \tabularnewline
\end{tabular}\end{ruledtabular}
\end{table}

Figure~\ref{fig1}(c) depicts the phonon dispersion of the BaNiIO$_6$ along the $\Gamma$-$M$-$K$-$\Gamma$-$A$-$L$-$H$-$A|L$-$M|H$-$K$ high-symmetry paths, which was calculated using the density-functional perturbation theory~\cite{add47,add48}. The phonon dispersion exhibits no imaginary modes, reflecting that it is dynamically stable. The thermodynamic stability of a 3 $\times$ 2 $\times$ 2 BaNiIO$_6$ superlattice was tested using AIMD simulation~\cite{add49} at 300 K for 3 ps in a Nos$\acute{\rm{e}}$-Hoover thermostat ensemble. The results are shown in Fig.~\ref{fig1}(d). During the simulation, the total energy was almost time-invariant, and the geometric configuration did not suffer any noticeable disturbance, indicating that BaNiIO$_6$ is stable at room temperature.

\begin{figure*}
\includegraphics[width=13.5cm]{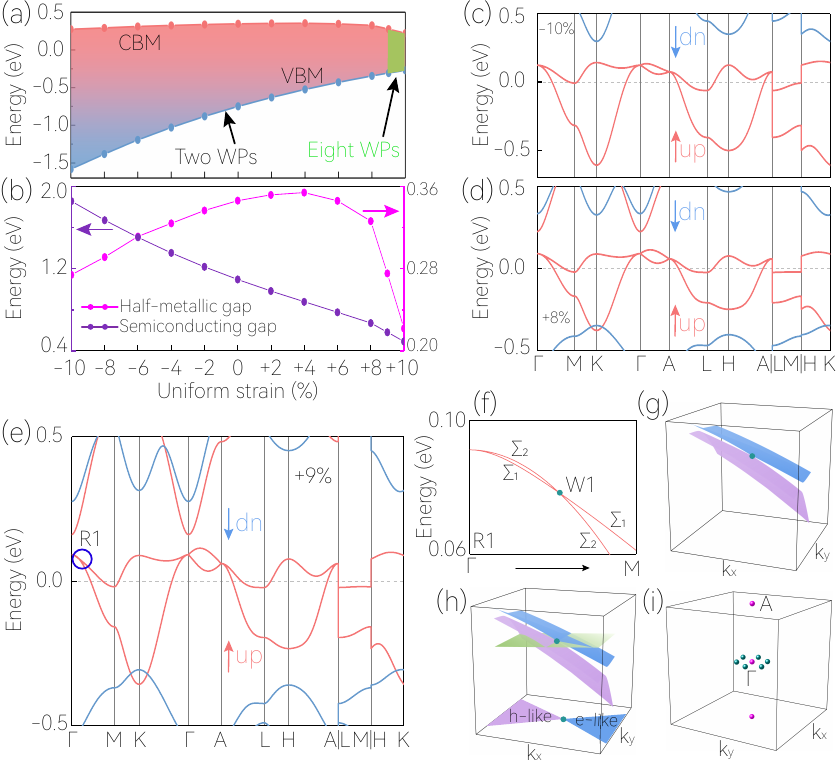}
\caption{(a) CBM and VBM as functions of uniform strains from -10\% to +10\%. The semiconducting gaps in the spin-down channel are highlighted by red and green backgrounds. The spin-polarized SP-WPs in the spin-up channel can be maintained from -10\% to +8\% uniform strains. However, eight spin-polarized WPs appear in the 3D BZ under +9\% and +10\% uniform strains. (b) Relationships between the uniform strains and the half-metallic/semiconducting gaps. Spin-polarized band structures of BaNiIO$_6$ under (c) -10\% and (d) +8\% uniform strains. (e) Spin-polarized band structure of BaNiIO$_6$ under +9\% uniform strain. A doubly degenerate crossing point (W1) appears along the $\Gamma$-$M$ path (in R1). (f) Enlarged spin-up bands of R1. Irreducible representations of the two bands around the W1 point on $\Gamma$-M path are also exhibited. (g) The 3D plot of the bands around the type-\uppercase\expandafter{\romannumeral2} WP W1 in R1. (h) Constant energy contour in the $k_x$-$k_y$ plane at the energy of the WP W1 in R1. The type-\uppercase\expandafter{\romannumeral2} dispersion hosts an electron-hole-like pocket Fermi surface. (i) Eight WPs (six type-\uppercase\expandafter{\romannumeral2} WPs and two type-\uppercase\expandafter{\romannumeral3} WPs) in the 3D BZ for BaNiIO$_6$ under +9\% uniform strain.
\label{fig3}}
\end{figure*}

\section{Robust half-metallic behavior}
Figure~\ref{fig2}(a) exhibits the calculated spin-polarized band structures and the density of states for BaNiIO$_6$ along the $\Gamma$-$M$-$K$-$\Gamma$-$A$-$L$-$H$-$A|L$-$M|H$-$K$ high-symmetry paths. Noticeably, the band structures have two spin directions (highlighted by blue and red colors). The spin-up bands overlap with the $E_F$, indicating metallic behavior; however, the spin-down bands exhibit semiconducting behaviors since the $E_F$ is located between the large gap ($\sim$1.09 eV) formed by the lowest conduction band (CB) and the highest valence band (VB) of the spin-down direction.

We calculated the spin polarization ratio ($P$) of BaNiIO$_6$ based on the density of states using the following formula~\cite{add50,add51}:
\begin{eqnarray}\label{1}
P=\frac{\mathrm{N} \uparrow\left(E_{F}\right)-\mathrm{N} \downarrow\left(E_{F}\right)}{\mathrm{N} \uparrow\left(E_{F}\right)+\mathrm{N} \downarrow\left(E_{F}\right)},
\end{eqnarray}
where $\mathrm{N}\uparrow(E_F)$ and $\mathrm{N}\downarrow(E_F)$ are the spin-up and spin-down electrons at the $E_F$, respectively. Since $\mathrm{N}\downarrow(E_F)=0$, BaNiIO$_6$ should have a full spin polarization around the $E_F$. Moreover, as shown in Figs.~\ref{fig3}(a) and S3(a), the half-metallic behavior (i.e., 100\% $P$) is maintained under different uniform strains (-10\% to +10\%) and different on-site Hubbard-Coulomb interactions ($U_{\rm{Ni}}$) for the Ni-$d$ orbital (0-6 eV).

As $U_{\rm{Ni}}$ changes from 0 eV to 6 eV, the semiconducting gap, which is the sum of the CB minimum (CBM) and the VB maximum (VBM) in the spin-down direction, increases from 1.09 eV to 2.56 eV (see Fig. S3(b)). The half-metallic gap, which is the smaller value between the absolute values of the CBM and VBM in the spin-down direction, can be viewed as an intuitive indicator of the robustness of the half-metallic behaviors. Figure S3(b) shows that the half-metallic gap~\cite{add52,add53} in the spin-down direction increases, attains a maximum value (0.825 eV), and then decreases as the $U_{\rm{Ni}}$ changes from 0 eV to 6 eV.

Figure~\ref{fig3}(b) depicts the relationship between the half-metallic and semiconducting gaps and the uniform strains. Figure~\ref{fig3}(b) shows that the semiconducting gap decreases from 1.856 eV to 0.491 eV as the uniform strain changes from -10\% to +10\%. However, the half-metallic gap continuously increases, attains a maximum value (0.354 eV), and then decreases as the uniform strain changes from -10\% to +10\%.

\section{Spin-polarized single pair of type-\uppercase\expandafter{\romannumeral3} WPs with $\lvert{\cal{C}}\rvert = 2$}
The spin-up bands around the $E_F$ are extremely clean. Specifically, only two doubly degenerate crossing points appear at the $A$ and $\Gamma$ TRIM points, and they are slightly above the $E_F$ (see the blue dots in Fig.~\ref{fig2}(a)). The inset of Fig.~\ref{fig2}(a) is a 3D Fermi surface at the $E_F$, demonstrating the appearance of the SP-WPs in 3D BZ. The density of states in Fig.~\ref{fig2}(a) and the orbital-resolved band structures in Fig. S4 demonstrate that the spin-up bands that form the SP-WPs are dominated by the Ni-$d$ and O-$p$ orbitals. Note that the SW-WPs can also be found in the spin-down channel (see Fig. S5 in SM~\cite{add46}), and they are also dominated by the by the Ni-$d$ and O-$p$ orbitals (see Fig. S6).

\begin{figure}
\includegraphics[width=8.5cm]{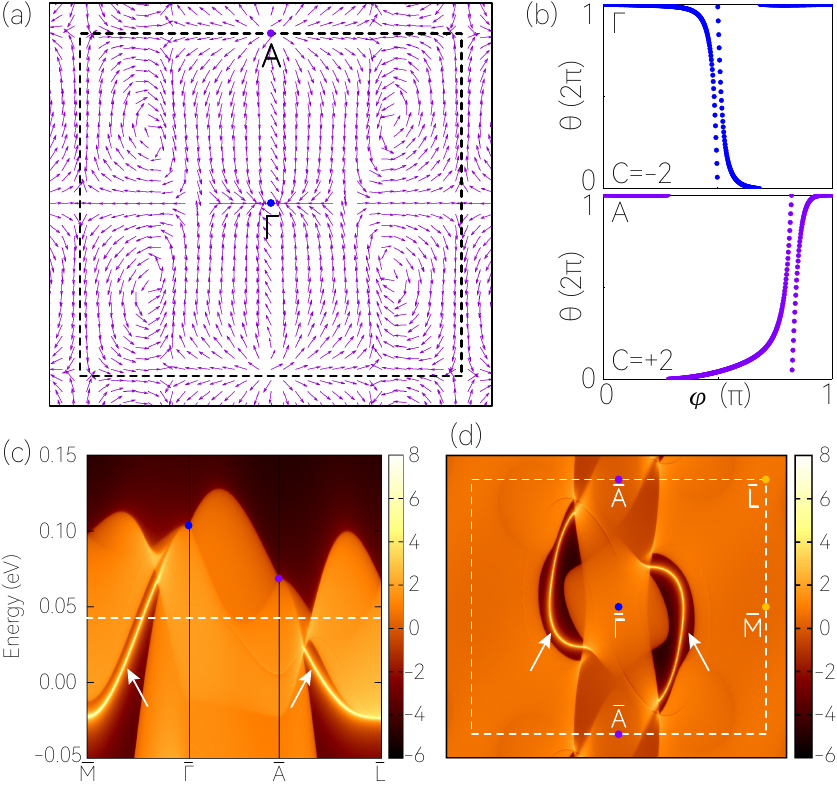}
\caption{(a) Distribution of the Berry curvature in the $k_y = 0$ plane. (b) Evolutions of the average position of the Wannier center for the WP with ${\cal{C}} = -2$ at $\Gamma$ and the WP with ${\cal{C}} = +2$ at $A$. (c) Projected spectrum of the (10$\overline{1}$0) surface states of BaNiIO$_6$ along the $\overline{M}$-$\overline{\Gamma}$-$\overline{A}$-$\overline{L}$ surface paths. (d) Energy slice for BaNiIO$_6$ corresponding to the (10$\overline{1}$0) surface at 0.045 eV (see the white dotted line in (c)). The two spin-polarized surface states that arise from the spin-polarized SP-WPs are marked by arrows.
\label{fig4}}
\end{figure}

Furthermore, as illustrated in Fig.~\ref{fig4}(b), the WP at $A$ has ${\cal{C}} = +2$, and the WP at $\Gamma$ has ${\cal{C}} = -2$ based on the evolutions of the average Wannier center positions~\cite{add54,add55}. Note that the SP-WPs at the $A$ and $\Gamma$ TRIM points with $\lvert{\cal{C}}\rvert = 2$ in the BaNiIO$_6$ system are strictly constrained by the no-go theorem~\cite{add30,add31}. Figure~\ref{fig4}(a) illustrates the corresponding distribution of the Berry curvature in the $k_x$-$k_z$ plane. The WP at $A$, which has ${\cal{C}} = +2$, acts as the ``source" point, while the WP at $\Gamma$, which has ${\cal{C}} = -2$, acts as a  ``sink" point, resulting in the BaNiIO$_6$ system having a neutral chiral charge.

When SOC is neglected, the WPs with $\lvert{\cal{C}}\rvert = 2$ at the $A$ and $\Gamma$ TRIM points are symmetry-enforced. Figure~\ref{fig2}(a) shows that the SP-WPs at the $A$ and $\Gamma$ TRIM points host a linear band dispersion along the $k_z$ direction (i.e., the $\Gamma$-A path). However, the SP-WPs host a quadratic band dispersion in the $k_x$-$k_y$ plane (see Fig.~\ref{fig2}(b) for the example of the 3D plot of the bands around the WP at the $A$ point in the $k_x$-$k_y$ plane). We display the 3D plot of the bands around the WP at the $A$ point in the $k_x$-$k_z$ plane as an example to clearly illustrate the type-\uppercase\expandafter{\romannumeral3} dispersion. The crossing bands possess a saddle-like dispersion. Figure~\ref{fig2}(d) shows the constant energy surface at the WP (at the $A$ point); there is a connection between two electron-like pockets.

When $\Gamma_3$ and $A_3$, the irreducible corepresentations of space group $P312$, are considered, the $\boldsymbol{k\cdot p}$ model can be easily obtained using the recently developed MagenticKP package~\cite{add55a,add55b}:
\begin{eqnarray}\label{1}
\begin{split}
H=\varepsilon+\omega_\Vert k_{\Vert}^{2}+\omega_zk_z^{2}+v_zk_z\sigma_3+ \\
((ak_+^{2}+bk_zk_+)\sigma_++h.c.),
\end{split}
\end{eqnarray}
where $\sigma_i(i=1, 2, 3)$ are three Pauli matrices, $\sigma_{\pm}=(\sigma_1\pm\sigma_2)/2$ and $\varepsilon, \omega_\Vert, \omega_z, v_z, a, b$ are real parameters, and $k_\Vert^{2}=k_x^2+k_y^2$. Note that $P312$ is a symmorphic space group, and the small co-groups of $\Gamma, A$ are isomorphic to each other; therefore, such a model applies to both $\Gamma_3$ and $A_3$. The dispersion on the $k_z=0$ or $\pi$ plane is $E(k_x,k_y)=\varepsilon+(\omega_\Vert\pm a)k_\Vert^2$. Thus, the Weyl cone for $\lvert\omega_\Vert\rvert>\lvert a\rvert$ can be over tilted, resulting in a type-\uppercase\expandafter{\romannumeral3} WP. Table~\ref{Tab2} shows the fitted parameters for $\lvert\omega_\Vert\rvert$ and $\lvert a\rvert$ of the two bands around the two WPs at $\Gamma$ and $A$, which confirm that the touching points at $\Gamma$ and $A$ are type-\uppercase\expandafter{\romannumeral3} WPs.

\begin{table}
\renewcommand\arraystretch{2}
\caption{\label{Tab2}The fitted parameters for $\lvert\omega_{\Vert}\rvert$ and $\lvert a\rvert$ of the two bands around the WPs at the $\Gamma$ and $A$ high-symmetry points.}
\begin{ruledtabular}
\begin{tabular}{ccc}
       High-symmetry points & $\lvert a\rvert$ (eV$\cdot\rm{{\r{A}}}^2)$ & $\lvert\omega_{\Vert}\rvert$ (eV$\cdot\rm{{\r{A}}}^2)$ \tabularnewline
       \hline
    $\Gamma$ & 0.34 & 0.91  \tabularnewline
    $A$ & 0.93 & 1.71       \tabularnewline
\end{tabular}\end{ruledtabular}
\end{table}

It is interesting to note the following: (1) the SP-WPs in the BaNiIO$_6$ system are nearly affected by the SOC. Table S2-S4 shows that BaNiIO$_6$ prefers the ferromagnetic state with the spin aligned along the [101] direction in the unit cell, 1 $\times$ 1 $\times$ 2, and 2 $\times$ 2 $\times$ 1 supercells, respectively, under SOC. Figure S7 shows the band structures with SOC. Compared to the band structures without SOC (see Fig.~\ref{fig2}(a)), the bands around the $E_F$ are almost unaffected by SOC. The SP-WPs at $A$ and $\Gamma$ are opened by SOC with a gap of less than 0.15 meV (Fig. S7 in SM~\cite{add46}), which is below the resolution of ARPES and significantly lower than the energy scale of room temperature (26 meV).

(2) Figure S3 shows that the spin-polarized SP-WPs are robust to the on-site Hubbard-Coulomb interactions for the Ni-$d$ orbital ($U_{\rm{Ni}}$ from 0 eV to 6 eV).

(3) The spin-polarized SP-WPs remain unchanged under uniform strains in the range of -10\% to +8\%. Figures~\ref{fig3}(c) and~\ref{fig3}(d) show examples of the spin-polarized band structures of BaNiIO$_6$ under -10\% and +8\% uniform strains, where the spin-polarized SP-WPs can be clearly observed. Interestingly, when we applied +9\% and +10\% uniform strains, the number of WPs in the BaNiIO$_6$ system increased from 2 to 8. The positions of the two type-III WPs (at $A$ and $\Gamma$) under ambient condition and uniform strains are shown in Fig. S8 (see SM~\cite{add46}). The spin-polarized band structures of BaNiIO$_6$ under +9\% uniform strain are shown in Fig.~\ref{fig3}(e) as a typical example. Apart from the two crossing points at $A$ and $\Gamma$, there is another doubly degenerate band crossing on the $\Gamma$-$M$ path (see Fig.~\ref{fig3}(f)). Note that the crossing point on the $\Gamma$-$M$ path has a type-\uppercase\expandafter{\romannumeral2} band dispersion~\cite{add56,add57,add58,add59,add60,add61,add62} and $\lvert{\cal{C}}\rvert=1$. The type-\uppercase\expandafter{\romannumeral2} dispersion hosts an electron-hole-like pocket Fermi surface (see Fig.~\ref{fig3}(h)) due to linearly tilted bands (see Fig.~\ref{fig3}(g)). As shown in Fig.~\ref{fig3}(i), there are six type-\uppercase\expandafter{\romannumeral2} WPs with $\lvert{\cal{C}}\rvert=1$ and two type-\uppercase\expandafter{\romannumeral3} WPs with $\lvert{\cal{C}}\rvert=2$, resulting in a total of 8 WPs in the 3D BZ.

When the direct sum of $\Sigma_1$ and $\Sigma_2$ irreducible corepresentations along the $\Gamma$-$M$ are considered (see Fig.~\ref{fig3}(f)), the $\boldsymbol{k\cdot p}$ model for the degenerate points can be written as
\begin{eqnarray}\label{1}
H=\varepsilon+c_1k_x+c_2k_x\sigma_3+((\alpha k_y+\beta k_z)\sigma_++h.c.),
\end{eqnarray}
where $\varepsilon, c_1, c_2$ are real parameters, and $\alpha, \beta$ are complex parameters. The band dispersion on the $\Gamma$-$M$ line ($k_x=k_y=0$) is $H=\varepsilon+c_1k_x+c_2k_x\sigma_3$. Notably, when $\lvert c_1\rvert>\lvert c_2\rvert$, the tilt term dominates the band dispersion along the $\Gamma$-$M$, forming a type-II WP. By fitting the parameters, we discovered that $\lvert c_1\rvert=0.21$ (eV$\cdot\rm{{\r{A}}})$ and $\lvert c_2\rvert=0.03$ (eV$\cdot\rm{{\r{A}}})$, confirming that the crossing point on the $\Gamma$-$M$ has a type-\uppercase\expandafter{\romannumeral2} band dispersion. The effects of SOC on the BaNiIO$_6$ system with +9\% and +10\% uniform strains are discussed in SM (see Fig. S9 and Fig. S10 in SM~\cite{add46}).

We would like to highlight that type-\uppercase\expandafter{\romannumeral2} and type-\uppercase\expandafter{\romannumeral3} WPs can exhibit many novel distinctive physical properties since most of the low-energy behaviors of systems in solids are usually determined by the geometry of the Fermi surface~\cite{add41,add63,add64}. The coexistence of type-\uppercase\expandafter{\romannumeral2} and type-\uppercase\expandafter{\romannumeral3} WPs in a single solid may induce unique physical properties.

(4) In our recent work~\cite{add29}, we proposed that the surface Fermi arc for SP-WPs (in a nonmagnetic spinless system) is extended and exhibits a noncontractible winding topology on the surface BZ torus. Figures~\ref{fig4}(c) and~\ref{fig4}(d) depict the (10$\overline{1}$0) surface states of BaNiIO$_6$ along the $\overline{M}$-$\overline{\Gamma}$-$\overline{A}$-$\overline{L}$ surface paths and a constant energy slice at 0.045 eV, respectively. In contrast to the non-spin-polarized surface Fermi arc~\cite{add65,add66,add67,add68,add69,add70} in nonmagnetic SP-WPs systems, the two surface Fermi arcs that arise from the spin-polarized SP-WPs in BaNiIO$_6$ are fully spin polarized. Note that the Fermi arcs induced by the two type-III WPs on the (10$\overline{1}$0) surface can also be found when +9\% or +10\% uniform strain is applied (see Figure S11 in SM~\cite{add46}).

(5) \textit{This study is an expansion of our previous study~\cite{add29}.} In our previous study, we stated that nonmagnetic spinless systems can host SP-WPs at two TRIM points and identified 32 candidate space groups (see Table 1 in Ref.~\cite{add29}) that host such a minimum number of WPs. In this paper, we mention that ferromagnetic half-metallic materials with negligible SOC (such as BaNiIO$_6$) are good platforms for realizing spin-polarized SP-WPs in their band structures. Since the ferromagnetic spin-up and spin-down bands have no symmetry connecting them, they can be considered separately when SOC is neglected. In other words, the bands for each spin direction can be viewed as spinless systems in principle. Hence, the proposed 32 candidate space groups in Ref.~\cite{add29} can be used to search for SP-WPs half-metals (with negligible SOC) in the future.
\\

\section{Summary and remarks}
We theoretically proposed a type-\uppercase\expandafter{\romannumeral3} SP-WPs half-metal for the first time using $P312$ type BaNiIO$_6$ as an example. Our first-principle calculations revealed that the spin-polarized single pair of type-\uppercase\expandafter{\romannumeral3} WPs are localized at the $\Gamma$ and $A$ TRIM points in the first BZ and strictly follow the no-go theorem. Moreover, the metallic state in the spin-up direction and the semiconducting state in the spin-down direction cause BaNiIO$_6$ to exhibit a half-metallic behavior with 100\% spin polarization. The SP-WPs with $\lvert{\cal{C}}\rvert = 2$ induce two Fermi arcs with extended surface states that wind around the surface BZ along the $k_z$ direction. The half-metallic state and the type-\uppercase\expandafter{\romannumeral3} SP-WPs in BaNiIO$_6$ are robust to the effect of uniform strains (from -10\% to +8\%) and on-site Hubbard-Coulomb interactions (from 0 eV to 6 eV). When +9\% or +10\% uniform strain was applied to the BaNiIO$_6$ system, the number of WPs increased from 2 to 8 in the 3D BZ. In such a case, there were eight WPs (six type-\uppercase\expandafter{\romannumeral2} WPs with $\lvert{\cal{C}}\rvert=1$ and two type-\uppercase\expandafter{\romannumeral3} WPs with $\lvert{\cal{C}}\rvert=2$) in the 3D BZ.

When SOC is neglected, the spin-up and spin-down channels in ferromagnetic half-metals are decoupled with a specific spin polarization axis, making it possible to view the bands for each spin direction as spinless systems. Note that our previous work~\cite{add29} has already guided the search and design of nonmagnetic spinless systems with SP-WPs, and it can also be used as a guide to design new SP-WPs ferromagnetic half-metals (with negligible SOC).

Our study undoubtedly expands the scope of SP-WPs materials, proposes the concept of type-\uppercase\expandafter{\romannumeral3} SP-WPs half-metals, and provides an example material, BaNiIO$_6$, for investigating their topological and spintronic properties.

Note that we are intrigued by a recent study~\cite{add44}. Li \textit{et al.}~\cite{add44} reported the experimental observation of type-\uppercase\expandafter{\romannumeral3} SP-WPs in nonmagnetic sonic crystals with space group No. 89 based on our previous work~\cite{add29}. Therefore, we hope that our work can provide new ideas for researchers in the field of spintronics and topology physics. The experimental confirmation of the type-\uppercase\expandafter{\romannumeral3} SP-WPs half-metals is imminent.

The investigation the topological signatures in half-metals is a hot research topic very recently. For example, one of the most known half-metals, the CrO$_2$, was theoretically demonstrated to host topological nodes~\cite{add70a}, as effectively shown in the probed experimental band structure~\cite{add70b}. Moreover, the theoretical prediction of Cr$_2$C as a 2D half-metal hosting spin-polarized WPs has been reported by Meng $et$ $al.$~\cite{add70c}. 

Finally, we would like to point out that $P6_322$ type BaNiIO$_6$ is also a dynamically stable ferromagnetic half-metal with eight WPs, formed by the crossings of the bands 62 and 63, in theory (see Fig. S12 in SM~\cite{add46}). The detailed results of the half-metallic and topological properties for the $P6_322$ type BaNiIO$_6$ will be presented in a separate paper.

\section{Acknowledgments}

This work was supported by the National Key R$\&$D Program of China (Grant No. 22022YFA1402600), the National Natural Science Foundation of China (Grant No. 12004028 and Grant No. 51801163), and the Natural Science Foundation of Chongqing (Grant No. CSTB2022NSCQ-MSX0283).

\end{document}